# Space Quantization of Light Transmission by Strong Coupling of Plasmonic Cavity Modes with Photosynthetic Complexes.

Itai Carmeli[1], Moshik Cohen[2], Omri Hieflero[3], Igal Liliach[4], Zeev Zalevsky[2], Vladimiro Mujica[5], Shachar Richeter[1]

[1]Department of Materials Science and Engineering, Faculty of Engineering & Center for Nano Science and Nanotechnology, Tel Aviv University, Tel-Aviv, 69978, Israel.

[2]Faculty of Engineering, Bar-Ilan University, Ramat-Gan 52900, Israel.

[3]Department of Chemistry and Biochemistry, Tel-Aviv University, Tel-Aviv 69978, Israel.

[4]Center for Nano Science and Nanotechnology, Tel-Aviv University, Tel-Aviv 69978, Israel.

[5]Department of Chemistry and Biochemistry, Arizona State University, Tempe 85287-1604, AZ.

The interaction between molecules and surface plasmons (SP) in defined geometries can lead to new light-mater hybrid states where light propagation is strongly influenced by molecular photon absorption[1–15] . This phenomenon has been observed in organic semiconductor optical microcavities [4,6,8,9,11–13], sub-wavelength hole arrays [1,5,7] and molecules on patterned surfaces. Their application range from lasing[16,17] LED's [18] to controlling chemical reactions[9,19] and are relevant in light harvesting[3,20,21]. The coupling between the electromagnetic field and molecular excitations may also lead to macroscopic extended coherent states characterized by an increase in temporal and spatial coherency[5,6,15–17] . In this respect, it is intriguing to explore the coherency of the hybrid system for molecules that possess highly efficient exciton energy transfer. Such a molecule, is the photosynthetic light harvesting complex photosystem I (PS I) which has an extended antenna system dedicated for efficient light harvesting. In this work, we demonstrate space quantization of light (SQOL) transmission through a single slit in free standing Au film coated with several layers of PS I. A self-assembly technique for multilayer fabrication is used, enabling fabrication of multilayers which leaves most of the hole vacant with only the surface and slit walls coated with molecules. When a broad band, non –coherent white light source excites the cavity, a strong SQOL is observed which is attributed to molecular photon absorption that induces coherency in the plasmon cavity modes. The SQOL is accompanied by a 13 fold enhancement in the extraordinary optical transmission (EOT) through the cavity. This work demonstrate the emergence of spatial coherency in a cavity strongly coupled to one of the most efficient energy transfer photosynthetic protein in nature and provides the path for engineering quantum electrodynamics by the vast diversity of electronic properties of biological macromolecules.

Photosynthetic complexes are evolutionary tuned to efficiently capture solar light. The initial steps of photosynthesis involve the absorption of photons by pigment molecules located in a protein antenna matrix which is followed by rapid and highly efficient funneling of exciton energy to power the reaction centers, in these specialized biological solar cells. Today there is compelling experimental evidence that long-lived electronic quantum coherence and entanglement play an important part in energy transfer processes in such systems even at physiological temperatures[22–24].

The integration of organic molecules with solid state materials has led to bio-photonic devices with remarkable properties, such as strong coupling with large Rabi splitting [1,3–5,7,9–11,13–15,18,19,25] and the appearance of collective coherent states demonstrating excitonic-polaron nature with coherence lengths of several microns[6].

Here, we report on the first successful attempt to merge an active photosynthetic bio-molecule into a quantum electro dynamic (QED) solid state device. We study EOT of a hybrid system composed of a PS I and a microscopic cavity drilled in free standing 200nm thick Au metal film (Fig.1). In general EOT is a phenomenon in which light transmission through sub wavelength apertures is greatly enhanced due to coupling of photons to surface plasmons at lower metallic interfaces. These plasmons than tunnel to the other side of the slit and re coupled to the radiating photons[26].

The photoactive reaction center PSI used in this study is a nanosize protein-chlorophyll complex which integrates 96 chlorophyll and 22 carotenoid pigment molecules in a helical protein membrane complex. It harvests photons with a quantum efficiency of ~1 and was found to be stable and photoactive when adsorbed in dry environment[21,27–29].

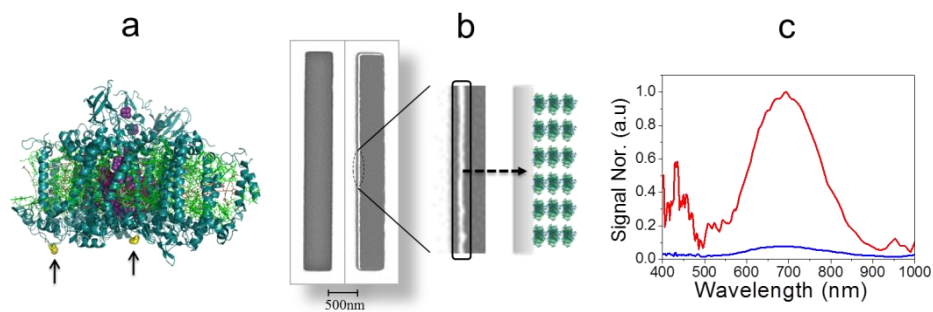

**Figure 1**: Coupling of PSI to free standing Au cavity. **a.** PSI structure. The cysteine mutations used to bind the PS I to the metal surface are depicted in yellow space fill sulfur atoms and indicated by arrows. **b.** left. SEM images of an uncoated 440×4500nm slit (left) and a slit which is adsorbed with 3 PS I layers (right). **b.** right. Zoom on the coated slit. The film corresponds to layer thickness of 33±4nm that is formed on the walls of the slit, leaving the majority of the slit vacant. **c.** Normalized transmission of light through 440×4500nm slit adsorbed with a multilayer composed of 3 PS I layers (red), in comparison with transmission through an uncoated slit (blue). Transmission of light through the coated slit is enhanced by a factor of 13 in comparison with the slit not coated with molecules.

The photosynthetic proteins self-assembled as monolayer or multilayer (SAM's) were derived from cyanobacterial membrane and adsorbed on the patterned Au metal films by formation of sulfide bond between the cysteine mutants of PSI and the metal surface (Fig.1a and method section)[27]. The advantage of this technique compared to spin coating, in which the entire hole is covered with a thick organic layer, is that here the surface of the metal and the slit circumference are covered with only a thin (few tenths of nm) of organized organic layer while leaving the majority of the slit void.

Transmission images and spectra recorded from slits with various dimensions (PSI-coated and uncoated) are sown in Figures 1c and 2. The spectral data shows a pronounced enhancement in EOT exhibiting 13 fold increase in light transmission for a slit coated with 3 PS I layers (Fig.1b,c). Transmission images (Fig.2) indicate that light is transmitted in a spatially smooth continuous manner in the case of bare (Fig.2a) or single-monolayer coated slit(fig.2b). Remarkably, when the surface of the slit is coated with three PS I layers, light emerge in circles pattern dividing the slit into several bright spots (Fig2.c-e). This collective phenomenon is clearly a combined result of molecular absorption and light propagation.

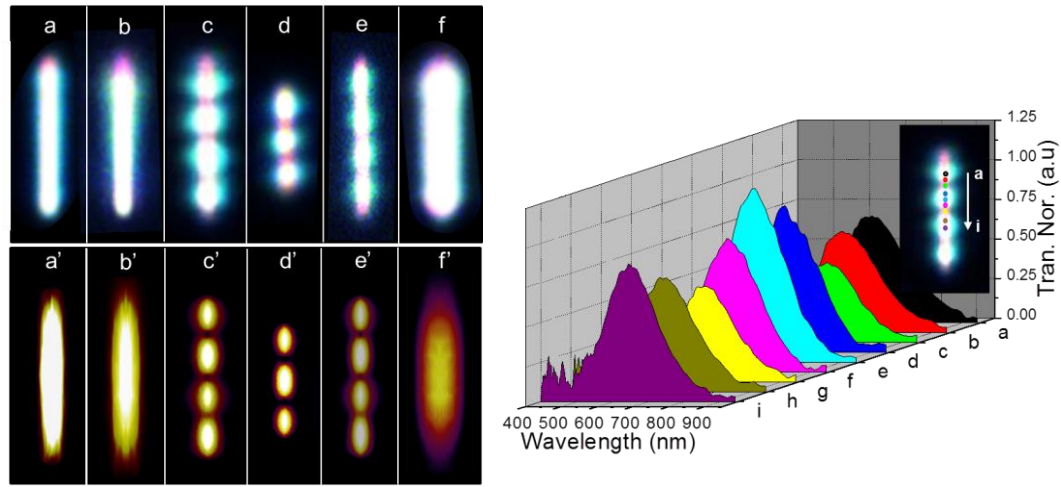

**Figure 2**: Light transmission through slits: experiment and simulations. Left: Transmission of white light through coated and uncoated slits (a-f) and the corresponding 3D simulation (a'-f'); **a.** an uncoated 440×4500nm slit **b.** 440×4500nm slit decorated with single PS I monolayer (**1PSI**) **c.** 440×4500nm slit coated with 3 PSI layers (**3PSI**). **d.** 440×3000nm, **3PSI**. (e) 220×4500nm, **3PSI**. (f) 900×4500nm, **3PSI**. Bright circles of light appear in slits coated with **3PSI** layers for slits width< 680nm. (a'-f') Simulation of transmitted light through the various slits.
Right: Spatial modulation of light transmission across the slit in Fig.2c. Inset: Scan direction is from point a to j. Light transmission intensity is modulated across the slit with maximum intensities at the center of the light circle and minimum at the nodes between two adjacent circles. Each spectrum was taken from an area cross section of ~80×80nm$^2$ of the CCD image.

The appearance of the pronounced SQOL does not depend on the width of the slit up to 680nm. However, at larger widths (see for example 900 nm Fig. 2f) the SQOL effect is lost indicating that light geometric resonances play an important role in this phenomenon. A representative spectral response of the spatially resolved light is shown in figure 2 right. From the spectra It is evident that light intensity across the slit length is modulated with minimum transmission at the nodes and maximum transmission in the center of the circles. A corresponding simulation performed by full 3D solution of Maxwell's equations shown in figure 2 (a'-f') was able to reproduce the main spectral features observed in the images (see theory section and SI for additional details).

Next, we investigate the SQOL with respect to the PSI absorption spectra (Fig. 3 right). Figure 3 left shows the transmission images taken under excitations in different bands: blue (420-480nm), green (510-550nm) and red (600-800nm). The images provide a direct evidence that a strong SQOL takes place only at wavelengths in which the PSI absorbs light. Interestingly, excitation by red and white light (4 circles) results in different quantization modes than excitation in the blue (6 circles). This observation (supported by simulations taken at center excitation frequency of each band, figure 3) indicates that the formation of the standing waves patterns is a result of the molecular excitation frequency coupled to the cavity modes.

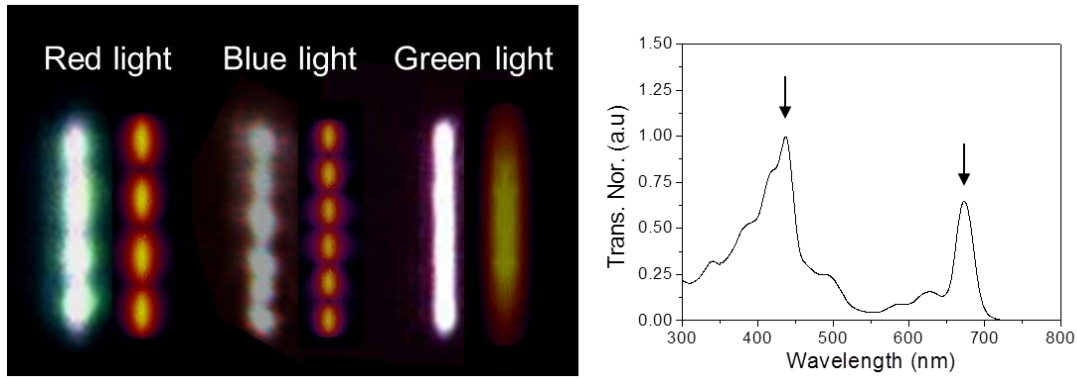

**Figure 3**: Dependence of transmission on PS I absorption bands. Left: the dependence of transmission on the wavelength of the transmitted light. Blue and red light are spatially modulated when transmitted through a 440×4500nm slit coated with three PS I layers, forming SQOL, while green light transmission shows legible spatial modulation. Simulations are depicted on the right side of each slit. Right: excitation spectra of the PS I in solution. Two distinct main absorption bands, in the blue (440nm) and in the red (680nm) are depicted by black arrows.

Spatial energetic distribution across slit 2d is presented in figure 4. The spectra show that EOT energy is spatially modulated across the slit.

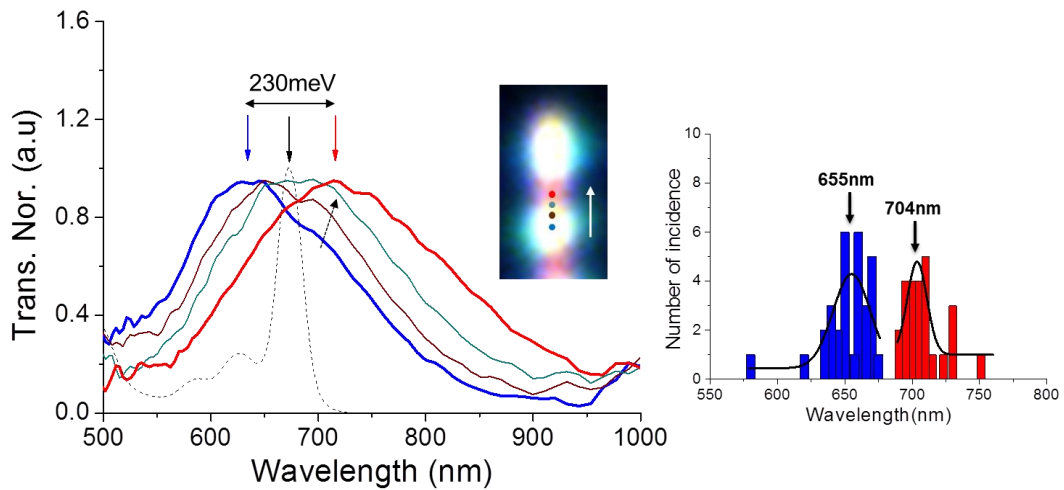

Figure 4: Spatially resolved transmission spectra of white light through a slit coated with 3 PS I layers. Left: Transmission spectra longitudinal scan across the 440×3000nm slit adsorbed with three PS I layers. Transmission is from the center of the white circle shown in the inset and represented by the blue curve, toward the node (red curve). The black doted curve is the absorption spectra of PS I in solution. The two transmission peaks energies pointed by the blue (636nm) and red (716nm) arrows are distributed evenly around the excitation energy of the PS I with an energetic separation of 230meV. Right: Distribution of peak position taken from the center of the white circle (blue histogram) and center of the pink band (red histogram) for 12 independent slits, fitted by Gaussian.

Apparently, the spectrum at the center of the white circle is composed of two contributions, one which originates from the center of the circle (636nm) and a shoulder (710nm) from the node. The energy of the two transmission peaks at 636nm and 716nm are distributed evenly on both sides of the excitation energy of the PS I at 680nm with a splitting of 230meV. The observation of energetic splitting is supported by statistical measurements taken for 12 individual slits (Fig.4, right).

The transmission of light through a hybrid PS I micro cavity is characterized by several distinct features. The first is the appearance of a strong and well-defined SQOL only in cases where the slits are coated with 3 PS I layers. SQOL takes place even for a broad band white light source excitation. There is an enhancement of a factor 13 in light transmission intensity of slits coated with PS I multilayer compared to uncoated slits. The enhancement is most prominent at the adsorption bands of the PSI. Excitation by higher photon energy (blue) results in appearance of six quantized circles while transmission of white and red light gives four quantized circles. The cavity displays only these specific modes, and no other modes were observed. In addition, there is a spatial distribution of transmission energy across the slit with a statistical average peak to peak energy difference of ~130meV.

The PS I coupled to the microcavity SP seems to operate as a molecular antenna that captures light efficiently and couples it effectively to the SP which in turn propagates the light excitation to the other side of the slit. This process results in enhanced EOT. The coupling induces coherency in the SP that is observed as SQOL in the transmission images and captured in our simulations. The effect scales linearly with the number of PS I layers, which seems to be consistent with our model because photon adsorption should increase with the thickness of the monolayer, that ranges from 10 nm to 30 nm for a single and a triple layer, respectively.

In addition a strong spectral correlation between the light absorbed by the PS I and the appearance of SQOL is found. Notice that the SQOL of the white light is comparable to that of red light due to the fact that the intensity of the light source is higher for longer wavelengths and the gold plasmon response is weaker for shorter wavelengths. The spatial energetic splitting across the slit of 130meV suggests Rabi splitting (RS) which occurs when there is strong coupling of molecular excitation with the electromagnetic field of the cavity. In that case[4,6,8,9,11–13] the excitation of the molecules split the SP energy into two components of higher and lower energies, corresponding to two SPP modes with forbidden crossing around the excitation energy of the molecule. The experimental technique used here allows the observation of spatial induced RS with a highly anisotropic refractive index, which has been previously predicted[14] but has not been observed. The Rabi splitting around the PSI excitation energy and the observation of SQOL only at molecular excitation bands indicates on the crucial role of the PSI in this phenomenon. The emerging picture is of a PS I antenna system that captures light efficiently at the excitation bands of the PS I and transfers it coherently to the SP to whom it is strongly coupled. The photon energy is then released at the other end of the slit and observed as spatial modulation of the transmission with enhanced coherency of the cavity. The SQOL effect observed here is specific for the PS I system and was not observed for any other molecules studied by us such as porphyrins, polypeptides and saturated monolayers of hydrocarbon.

## Methods

The hybrid system was fabricated by first forming a freestanding Au film on a TEM grid (see illustration in the SI). This was achieved by evaporating 200nm Au on a NaCl crystal (Agar Scientific). The rate of evaporation was 0.7A/sec for the first 20nm and then increased to 2A/sec. After evaporation, the crystal was cut to ~3mm$^2$ pieces and placed in a petri dish of deionized water, until the gold leaf was detached from the crystal and floated on the water surface. The gold leaf was then mounted on an Au TEM grid (SPI supply) dried with cleanroom paper and placed for 2 minutes on a hot plate (60$^0$C) for further drying. The TEM grid was then washed in acetone and isopropanol and dried in air. In the second step slits of different dimensions were drilled with a focused Ion Beam (FIB). Adsorption of PS I monolayer was achieved by first cleaning the Au grid by Ar and O2 (50%:50%) plasma. The grids were than washed in ETOH for 2 minutes, dried and placed in a concentrated solution of PS I for 2 hours and washed in buffer and deionized water. Multilayers were fabricated by crosslinking of successive PS I layers. This technique insures that the PS I multilayer is covalently bound to the surface in an oriented fashion. An evident advantage of this technique unlike previously reported procedures of spin coating is that the organic layers here, covers the surface of the metal and the slit circumference with only a thin (few tenths of nm) organized layer leaving the majority of the slit void (Fig.1) were as in spin coating molecules fill the hole interior and they are unorganized.

## Experimental

In our setup, the sample (Au leaf on a TEM grid) is placed on top of a 2mm hole in plastic microscope slide and attached to the sides of the hole. The microscope slide is placed on the stage and centered at the beam. Light is provided by a 100W quartz halogen lamp at the base of the microscope (field diaphragm). Light enters the condenser and after passing through the sample the light enters a 100X magnifying objective which gives an image to the CCD. The CCD is capable of giving the spectral response at any given point of the image with a spatial resolution of 32nm and wavelength spectral resolution FWHM of 6nm. The spectral response is given by dividing the transmission spectra of the light passing through the sample with a blank reference (same illumination condition with absence of a sample). For testing wavelength effect on transmission color filters were used. The filters were placed in the beam path before it enters the condenser. The filters used transmit light, in the blue (420-480nm), green (510-550nm) and red (600-800nm).

## Simulation

The theoretical analysis was carried out using direct 3D solution of Maxwell's equations. The numerical solution is obtained via a finite element method (FEM) based on a commercial software package, HFSS v.15.2, which is broadly used for analysis of plasmonic nanostructures [30]. A complex dielectric function ε(ω) = ε$_1$(ω)+iε$_2$(ω) was used for modeling the PS_I complex. The imaginary part of the dielectric function was extracted from experimentally obtained frequency dependent absorption[31] coefficient using $\varepsilon_2(\omega) = \frac{\check{n}c}{\omega}\alpha_{abs}(\omega)$ where $\check{n}$ is the refraction index, c is the speed of light in free space, $\omega$ is the angular frequency and $\alpha_{abs}$ is the absorption coefficient[32]. The average real part of the dielectric function of PS_I, $\varepsilon_1(\omega)$, was obtained from[33]. The macroscopic optical properties of Au were obtained from ref [34]. Similar to the experimental setup, the structure was illuminated from below with non-polarized unfocused light which covers the entire slit (for additional details see Methods section and SI). The incident photons are converted to surface plasmons at the bottom side of the slit, channel through to the upper side of the slit and re - coupled to radiating photons. This phenomenon is referred to extraordinary optical transmission [26], as the width of the metallic slits is smaller than the wavelength of incident light.